\documentclass[aps,prc,preprint,groupedaddress,floatfix,showpacs,twoside]{revtex4}
\usepackage{graphicx}% Include figure files
\usepackage{dcolumn}% Align table columns on decimal point
\usepackage{bm}% bold math
\begin{document}

\setcounter{page}{1}
\pagestyle{myheadings} \oddsidemargin
4mm\evensidemargin -4mm

\begin{center}
{\bf Surface Boiling - a New Type of Instability of Highly Excited Atomic Nuclei\\}
\bigskip
\bigskip
\bigskip

{J. T\~oke and W.U. Schr\"oder\\}
{\it Departments of Chemistry and Physics\\
University of Rochester, Rochester, New York 14627}

%\date{\today}
\end{center}
\bigskip
\bigskip
\centerline{ABSTRACT}
\begin{quotation}
The evolution of the nuclear matter density distribution with excitation energy is studied within the framework
of a finite-range interacting Fermi gas model and microcanonical thermodynamics in Thomas-Fermi approximation. It is found that with increasing excitation energy, both infinite and finite systems become unstable against infinitesimal matter density fluctuations, albeit in different ways. In modeling, this instability reveals itself via an apparent negative heat capacity of the system and is seen to result in the volume boiling in the case of infinite matter and surface boiling in the case of finite systems. The latter phenomenon of surface boiling is unique to small systems and it appears to provide a natural explanation for the observed saturation-like patterns in what is commonly termed {\it caloric curves} and what represents functional dependence of nuclear temperature on the excitation energy.

\end{quotation}

%\pacs{21.65+f, 21.60.Ev, 25.70.Pq}
\newpage

\section{Introduction}

Understanding the limits of thermodynamical stability of excited nuclear systems has been a focus of numerous theoretical and experimental studies \cite{muller_serot,natowitz,shlomo_kolomietz} from the dawn of nuclear science. While experimental studies invariably involve finite nuclear systems formed in the course of nuclear reactions and thus not subjected to external confinement and the related pressure, theoretical studies are overwhelmingly concentrated on instabilities in bulk nuclear matter kept under controlled conditions. This then results, e.g., in phase separation and spinodal phenomena, the relevance of which for finite nuclear systems is far from obvious. The reason here is that it is not possible to bring the bulk matter in finite nuclei to the state showing in theoretical analysis the kind of instability of interest. Unlike in theoretical modeling, it is not possible to control the volume, pressure, or the temperature of actual nuclei and the best one can hope for is that by solely feeding more and more (excitation) energy into the system one will arrive at a point where the system becomes thermodynamically unstable, in addition to being trivially unstable against statistical particle evaporation, fragmentation, gamma, and beta decays.

To discover such instabilities in theoretical modeling of nuclear systems, one must then keep increasing the excitation energy, the only controlling parameter available in experiments, in infinitesimally small steps and checking if a metastable equilibrium is possible for a consecutive value of the excitation energy. The system is considered at metastable thermodynamical equilibrium when it may decay only as a result of {\it finite} statistical fluctuations in parameters describing the system as a whole. The onset of instability is in such a case signaled by the loss by the system of immunity against {\it infinitesimally small} fluctuations in one or more parameters, i.e., where such fluctuations no longer would give rise to restoring driving forces, but rather to disruptive forces. Mathematically, such instabilities reveal themselves in modeling through the appearance of negative compressibility, negative heat capacity, or negative derivative of chemical potential with respect to concentration \cite{lattimer_ravenhall,barranco_buchler,muller_serot}.

The present work is part of a continued effort
\cite{toke-swiatecki,toke_quantum,toke_surface,toke_surfentr,toke_retard,toke_box,
toke_unified,toke_signatures,sobotka_mono,sobotka_2006,sobotka_iso} to construct a (microcanonical)
thermodynamical framework for understanding phenomena of apparently statistical nature observed
in highly excited nuclear systems produced in the course of heavy-ion collisions. More specifically, it focuses on the
negative heat capacity observed \cite{toke_retard} for a thermally expanding bulk nuclear matter considered within the interacting Fermi gas model within Thomas-Fermi approximation. Now, for the first time, the onset of this negative heat capacity is identified as the boiling point for bulk nuclear matter, however, occurring at conditions substantially different from those typically attributed to boiling. Furthermore, model calculations performed for finite systems using finite-range interacting Fermi gas model combined with Thomas-Fermi approximation demonstrate that, on the excitation energy scale, much before the bulk matter would come to boiling, the surface matter starts boiling off, not allowing one to reach temperatures in excess of this surface boiling-point temperature. The phenomenon of surface boiling, discovered here via theoretical modeling, appears unique to small systems interacting via finite-range forces, such as is the case of atomic nuclei. It finds a solid experimental confirmation in the form of caloric curves "saturating" at temperatures of several MeV \cite{natowitz}, expected for surface boiling.

The present paper is structured as follows. In Section \ref{section_theory} the adopted theoretical formalism is presented based on finite-range interacting Fermi gas model and Thomas-Fermi approximation in conjunction with microcanonical statistical thermodynamics. In Section \ref{section_bulkboiling}, the boiling instability in bulk matter at zero pressure is revisited with clear demonstration of its nature. In Section \ref{section_surfaceboiling}, the evolution of the density profile of finite droplets of nuclear matter with increasing excitation energy is studied revealing the onset of instability against the infinitesimal fluctuations in the local matter density profiles, where such infinitesimal fluctuations are seen to give rise not to restoring forces but to effective driving forces amplifying the fluctuations to the point where parts of the surface matter separate from the system in what constitutes surface boiling.
The conclusions are presented in Section \ref{section_summary}

\section{Theoretical framework}
\label{section_theory}
A phenomenological model has been proposed and used earlier \cite{toke-swiatecki,toke_surface,toke_surfentr,toke_retard,toke_box,toke_unified,toke_signatures} to allow treatment of excited nuclei as droplets of unconfined Fermi gas/liquid in an (approximate) microcanonical equilibrium, where all allowed microstates are assumed to be equally probable. Since a true microcanonical equilibrium is not possible in a system excited in excess of particle binding energy, one considers the excited nuclear system to be bound in phase space by a hypersurface of transition states which can be reached by way of finite fluctuations and which act as effective doorways to decay channels. Examples of such transition states are states where at least one particle is in continuum at or in excess (for protons) of the Coulomb barrier or states at fission or fragmentation saddle configurations. The model is valid only in the domain of excitation energies, where such a hypersurface exists, i.e., where the system can be in a state of transient metastability, justifying the notion of an approximate microcanonical equilibrium. The micro-states are considered allowed when their energy is equal to the assumed model energy of interest and when they obey other conservation rules. Under such assumptions, common to all statistical models of excited nuclei, the probability for the system to reside in any particular macrostate or configuration is proportional to the number of microstates such configuration allows to visit, which is called here the configuration partition function $Z_{config}$ and logarithm of which is Boltzmann's configuration entropy, $S_{config}=lnZ_{config}$. Note that Boltzmann's $little-k$ is unity according to an adopted convention, where nuclear temperature is measured in units of MeV. The term configuration refers to a macroscopically distinct state of the system characterized by the matter distribution and, possibly, collective velocity field, such as, e.g., the one corresponding to collective rotation, vibration, or hypothetical self-similar expansion \cite{eesm}. In the present study, given its goals, the collective velocity field is assumed to be absent, securing highest entropy for any macroscopic matter distribution.

Accordingly to the notion of Boltzmann's entropy, the probability of finding the system in a given configuration is given by
\begin{equation}
P_{config} = {Z_{config}\over Z_{system}}={e^{S_{config}}\over e^{S_{system}}},
\label{eq_pconfig}
\end{equation}
\noindent where $Z_{system}$ and $S_{system}$ are the system partition function and entropy, respectively. The former can be expressed as
\begin{equation}
Z_{system} = \Sigma_i Z^i_{config},
\label{eq_zsystem}
\end{equation}
\noindent where the sum extends over all possible configurations. Since the number of such configurations is excessively large, Eq.~\ref{eq_zsystem} as a whole is impractical and one is always compelled to limit the sum on its R.H.S. to a manageably small number of configurations of interest, expected to be dominant. While often it is sufficient to consider only the most likely configuration out of the very many configurations described by a suitable parametrization of the matter distribution, one must keep in mind that potentially interesting phenomena are the ones where more than one configuration must be explicitly considered.

In view of the above, the important quantity to evaluate is the configuration (Boltzmann's) entropy, which the present formalism approximates via the zero-temperature interacting Fermi-gas model equation as
\begin{equation}
S_{config}=2\sqrt{a_{config}(E-E_{config})},
\label{eq_sconfig}
\end{equation}
\noindent where $E$ is the system energy and $a_{config}$ and $E_{config}$ are, respectively, the level density parameter and zero-temperature energy for the given configuration.

Equation \ref{eq_sconfig} is the base equation of the model, allowing one to evaluate $S_{config}$ for any configuration of interest characterized solely by the configuration matter density distribution $\rho_{config}(\vec{r})$. Such an evaluation involves evaluating separately $a_{config}$ and $E_{config}$. The former is done using Thomas-Fermi approximation \cite{toke-swiatecki} and the latter is done by integrating over volume the energy density given by a suitable equation of state, with a folding provision for mocking up the effects of the finite range of nucleon-nucleon interaction. For $a_{config}$ one writes \cite{toke-swiatecki}
\begin{equation}
a_{config} = \alpha_o\rho_o^{2/3}\int\int\int \rho^{1/3}(\vec{r})d\vec{r},
\label{eq_aconfig}
\end{equation}
\noindent where $\alpha_o$ expresses the value of the level density parameter per nucleon at normal matter density $\rho_o$.

The zero-temperature energy of a given configuration $E_{config}$ is taken as consisting of a potential (interaction) energy part $E^{EOS}_{int}$ and a kinetic energy part $E_{Pauli}$ arising from the action of the Pauli exclusion principle, i.e.,

\begin{equation}
E_{config}=E^{EOS}_{int}+E_{Pauli}.
\label{eq_econfig}
\end{equation}

The interaction energy is here calculated by folding a standard Skyrme-type EOS interaction energy density $\epsilon^{EOS}_{int}(\rho)$ with a Gaussian folding function with a folding length $\lambda$ adjusted so as to approximately reproduce the experimental surface diffuseness of finite droplets of nuclear matter, i.e.

\begin{equation}
E^{EOS}_{int}= R_{Gauss}\int\int\int\int\int\int \epsilon^{EOS}_{int}(\rho(\vec{r}-\vec{r'}))e^{-{(\vec{r}-\vec{r'})^2\over 2\lambda^2}}d\vec{r}d\vec{r'},
\label{eq_efold}
\end{equation}

\noindent where $R_{Gauss}$ is the normalization factor for the folding Gaussian. Note that the use of a folding integral of Eq.~\ref{eq_efold} makes the EOS non-local, such that the values of intensive thermodynamical parameters in a given location depend on the matter density distribution in neighboring locations. This results in the same problems, albeit on a smaller scale, that thermodynamics encounters when facing Coulomb and/or gravitational forces.

The Pauli energy was calculated from a Fermi-gas model expression

\begin{equation}
E_{Pauli} = {3\over 5}E^{Fermi}_o \rho_o^{-2/3}\int\int\int \rho^{5/3}(\vec r)d\vec r,
\label{eq_epauli}
\end{equation}

\noindent where $E^{Fermi}_o$ denotes the Fermi energy at normal matter density $\rho=\rho_o$, characteristic of the EOS adopted.

For the equation of state, the present study adopted the standard form consistent with Skyrme-type nucleon-nucleon interaction, which implies the interaction energy density (appearing in Eq.~\ref{eq_efold}) in the form
\begin{equation}
\epsilon^{EOS}_{int}(\rho)=\rho[a({\rho\over \rho_o}) + {b\over \sigma+1}({\rho\over \rho_o})^\sigma]
\label{eq_epsilon}
\end{equation}

The values of the parameters $a$, $b$ and $\sigma$ in Eq.~\ref{eq_epsilon} are determined by the requirements for the binding energy, matter density, and the incompressibility modulus to have prescribed values. The values chosen in this study of $a$=-62.43 MeV, $b$=70.75MeV, and $\sigma$ = 2.0 imply a normal density of $\rho_o=0.168fm^{-3}$, binding energy per nucleon at normal density of $\epsilon_{EOS}/\rho_o$=-16MeV, the incompressibility modulus of $K=220MeV$, and Fermi energy at normal density of $E^{Fermi}_o$=38.11MeV.

In the model calculations for uniformly distributed matter, the finite range of interaction is of no consequence and the configuration energy can be written simply as
\begin{equation}
E_{config}(\rho) = V\epsilon_{EOS}(\rho)
\label{eq_econfiguniform}
\end{equation}

\noindent where V is the system volume and $\epsilon_{EOS}(\rho)$ is the full configuration energy density (including interaction and Pauli energies) given by the equation of state as a function of matter density $\rho$.

The (microcanonical) temperature $T$, pressure $p$, and chemical potential $\mu$ were evaluated using the following model expressions

\begin{equation}
T = 1/({\partial S\over\partial E^*})_{V,N}=\sqrt{E^* - E_{config}\over a_{config}}\;\;,
\label{eq_t}
\end{equation}

\begin{equation}
p(\rho, T) = T({\partial S\over \partial V})_{E^*,N}=\rho_o[{\rho^2\over \rho_o}{d\bar{\epsilon}_{EOS}\over d\rho}+
{2\over 3}\alpha_o({\rho\over \rho_o})^{1/3}T^2]\;\; , and
\label{eq_p}
\end{equation}

\begin{equation}
\mu (\rho,T) = -T({\partial S\over \partial N})_{V,E^*}={\rho\over \rho_o}[\bar{\epsilon}_{EOS}+\rho_o{d\bar{\epsilon}_{EOS}\over d\rho}]
-{1\over 3}\alpha_o({\rho\over \rho_o})^{-2/3}T^2.
\label{eq_mu}
\end{equation}

In Eqs. \ref{eq_t}-\ref{eq_mu}, N represents the number of nucleons, $\alpha_o$ represents the value of the {\it little a} parameter per nucleon at normal density, and $\bar{\epsilon}_{EOS}$ represents configuration energy per nucleon.

It is perhaps amusing to note that according to Eq.~\ref{eq_mu}, at $T=0$ and at equilibrium density $\rho=\rho_o$, the chemical potential $\mu$ is here equal to the average (configuration) energy per nucleon $\bar{\epsilon}_{EOS}$, same as stated in the Hugenholtz-Van Hove theorem for a much more strict treatment of nuclear interactions than ours.

With a formalism set up for evaluating entropies for configurations of interest one has a sound thermodynamical framework for understanding a variety of statistical phenomena occurring in highly excited nuclear systems. In particular, within this framework, the decay rates into various decay channels are related to configuration entropies at the transition-states for the decay channels of interest, i.e. states at the transition-state hypersurface confining the model system. On the other hand, the (quasi-) equilibrium properties of the system in its microcanonical metastable state can be  inferred from finding the configuration of maximum entropy among the ones deemed to be relevant.
The latter kind of modeling, pursued in the present study, includes in a natural way researching the limits on metastability of excited nuclear systems, where no maximum entropy is found within a suitably parameterized space of configurations under consideration.

Note that the above formalism makes a number of simplifying approximations, such as neglecting the Coulomb forces, setting the effective nucleonic mass to the free nucleon mass, neglecting iso-spin effects, using zero-temperature Fermi-gas model expressions, etc. Also, the role of quantum effects on the Pauli energy \cite{toke_quantum} of finite systems is here neglected. These approximations may be dropped when issues other than studied in this work are to be addressed.

\section{Boiling instability in bulk nuclear matter}
\label{section_bulkboiling}

\begin{figure}
\includegraphics [width=8cm ]{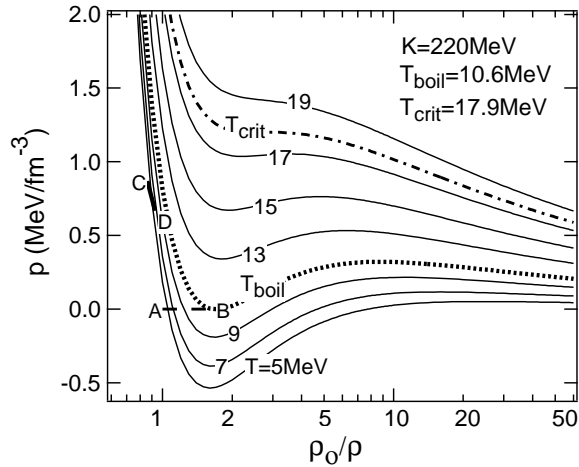}
\caption{Isotherms for the model matter. The isotherm corresponding to zero-pressure boiling temperature is shown in dotted line and the critical isotherm is shown in dash-dotted line. The adiabatic trajectory for a hypothetical infinite system is shown in dashes (line AB), while such for the bulk of a finite (A=100) system is shown in bold solid line (line CD).}
\label{fig_1}
\end{figure}

General behavior of uniform Fermi matter can be well understood from the appearance of isotherms in the familiar Wan der Waals type plots. These are illustrated in Fig.~\ref{fig_1} for the bulk model matter with Skyrme-type EOS with a compressibility constant of K=220 MeV. The isotherms are seen to feature spinodal domains of negative compressibility, which have been often discussed in literature in the context of general dynamical instability of the matter. The spinodal domains are generally inaccessible to experiment, as it is not possible to bring the system as a whole to uniform density, temperature, and pressure in these domains. Rather, the presence of these domains is an indication for onset of boiling (higher densities) or condensation (lower densities) instabilities as system parameters are varied, and thus of the possibility of phase separation and hypothetical phase coexistence, were it possible to transiently control pressure, volume, and temperature. The isotherms illustrate also the presence of a critical point, where the two (boiling and condensation) limits of the spinodal domain merge making the two phases identical.

One must recognize, however, that while the isotherm plots in Fig.~1 are quite instructive, they are largely of academic value in the case of nuclear matter. This is so because it is not possible to bring the nuclear matter to conditions spanned by these plots, as there is no meaningful way to control any of the volume, pressure, and temperature of real nuclear systems, while at the same time ensuring their bulk uniformity. What is experimentally accessible in Fig.~1, is a  mini-domain expressing the evolution of the bulk (uniform) matter density with excitation energy. For any definite system, whether infinite or finite, the accessible domain is strictly one-dimensional. For example, for hypothetical infinitely large systems this domain degenerates into a short section of a line (shown in Fig.~1 in dashes) at zero pressure, connecting the point A at normal density (p=0, T=0, $\rho=\rho_o$) and ending at the boiling point B at $p=0, T=T_{boil}=10.8 MeV$ and $\rho\approx 0.6\rho_o$. For the bulk (excluding the non-uniform surface domain) of a finite, A=100, system studied further below, the accessible domain degenerates into a short line CD (at $\rho>\rho_o$) shown schematically in Fig.~\ref{fig_1} in bold.

As noted above, much of the model space covered by Fig.~\ref{fig_1} is experimentally inaccessible to nuclear science. Yet, the very fact that the accessible domain is so limited and ends (with increasing excitation energy) always on a (boiling) spinodal, is indicative of the presence of a limiting excitation energy real nuclear systems residing in vacuum, i.e., at zero external pressure are able to equilibrate without becoming unstable. And it is the latter quantity that can be measured and thus compared to the model predictions.

\begin{figure}
\includegraphics [width=8cm ]{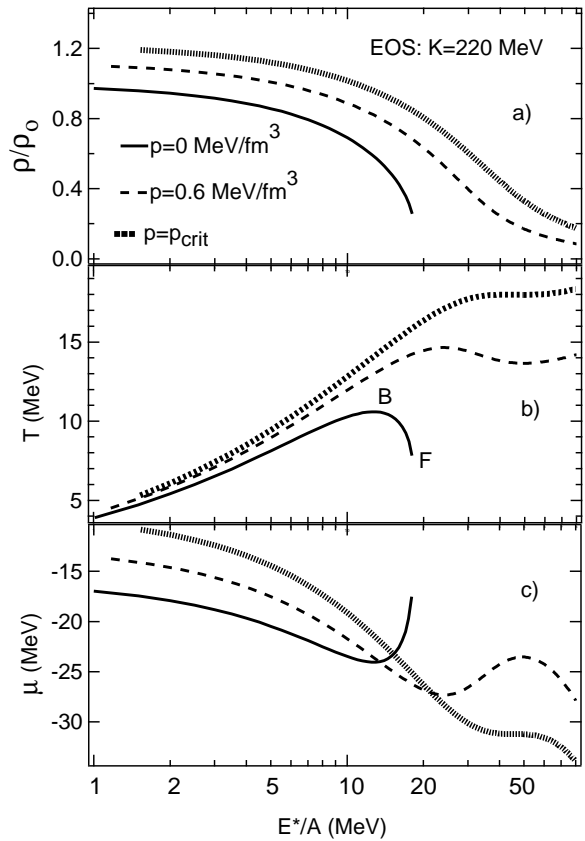}
\caption{Evolution of the equilibrium density (top panel), temperature (middle panel), and chemical potential (bottom panel) with excitation energy per nucleon for uniform Fermi matter with compressibility constant of K=220 MeV. The three lines in each panel are, respectively, for zero pressure (solid line), p=0.6 MeV/fm$^3$ (dashes), and the critical pressure $p_{crit}$ (dotted line).}
\label{fig_2}
\end{figure}

The response of the bulk uniform nuclear matter to the excitation energy is illustrated in Fig.~\ref{fig_2}. Here, for any given excitation energy per nucleon, the matter density was varied so as to produce any of the three select pressures - the natural zero pressure, critical pressure, and an intermediate pressure. The latter two are of academic interest only as they require an external containment manostat. Note that, consistent with the First Law of Thermodynamics, the requirement of zero pressure secures maximum (configuration) entropy $S_{config}$ for a given excitation energy and thus implies that the system is microcanonical (within the model constraint of uniformity).  Subsequently, (microcanonical) temperature $T$, pressure $p$, and chemical potential $\mu$ were determined for the equilibrium uniform configuration using the Fermi gas model expressions of Eqs.~\ref{eq_t}, \ref{eq_p}, and \ref{eq_mu}.

As seen in Fig.~\ref{fig_2}, the equilibrium density decreases monotonically with increasing excitation energy for all three values of pressure studied. On the other hand, temperature reaches maximum value and the chemical potential reaches a minimum value at $E^*/A$=17 MeV (zero pressure) and then they begin dropping (temperature) and increasing (chemical potential) with increasing $E^*$. This kind of behavior is indicative of the onset of thermodynamical instability at around $E^*/A$=17 MeV. In this domain, acquisition by any small part of the system of an infinitesimally small amount of excitation energy from neighboring parts via fluctuations would lead to a lowering of temperature in this part, which will then cause driving even more heat to it from hotter neighbors and a further lowering of temperature, until that part leaves the system. The latter follows from the fact that the zero-pressure curve ends at a point F ($E/A_F\approx 18 MeV$), where the entropy function no longer has a maximum as a function of the matter density $\rho$. Thus, any portion of the matter that reaches point F, will expand indefinitely on its own. Note, that it follows from the above, that the boiled off matter may be expected to be colder than the liquid residue, a fact that should be verifiable experimentally. The latent heat for boiling off is $E/A_F-E/A_B\approx 5.5 MeV$ per nucleon.

A narration perhaps better suited for nuclear systems, where processes are mediated by nucleon transport, is based on the behavior of the chemical potential and, specifically, on the fact that in a spinodal domain this potential features (destabilizing) negative first derivative with respect to concentration. In such a domain, when in one part of the system the concentration (matter density) decreases via fluctuations, that part increases its chemical potential and thus begins now feeding flux to neighboring parts and thus increasing its chemical potential even further.

In fact, the system would never arrive at a configuration showing nominally negative heat capacity or negative chemical susceptibility but would rather boil off "offending" parts of the matter while tending toward a metastable equilibrium. This (preequilibrium) boiling allows the remainder to shed the excess excitation energy, leaving a metastable residue at an excitation energy per nucleon equal to that at the boiling point ($E/A_B\approx 12.5 MeV$). Fig.~\ref{fig_2} illustrates additionally caloric curves calculated for hypothetical uniform system subjected to two different non-zero external pressures. As seen in this figure, at (hypothetical) critical pressure, the system stays always stable and uniform, while at (hypothetical) intermediate pressure, the system would find stability in a two-phase configuration.

According to the above analysis, the condition for the boiling can be expressed via a set of two equations, such as

\begin{equation}
({\partial S\over \partial \rho})_E=0\;\; and
\label{eq:boil_a}
\end{equation}
\begin{equation}
{\partial^2 S\over \partial E^2}=0.
\label{eq:boil_b}
\end{equation}

Note that the presence of a domain of negative heat capacity can be inferred already from the appearance of the isotherms seen in Fig.~\ref{fig_1}. Clearly, by following the zero-pressure line in the direction of decreasing matter density, representing the system trajectory as a function of excitation energy, one would first cross consecutive isotherms with increasing temperature labels, but then, after reaching the boiling-point temperature, the system trajectory would cross consecutive isotherms with ever decreasing temperature labels.

\begin{figure}
\includegraphics [width=10cm ]{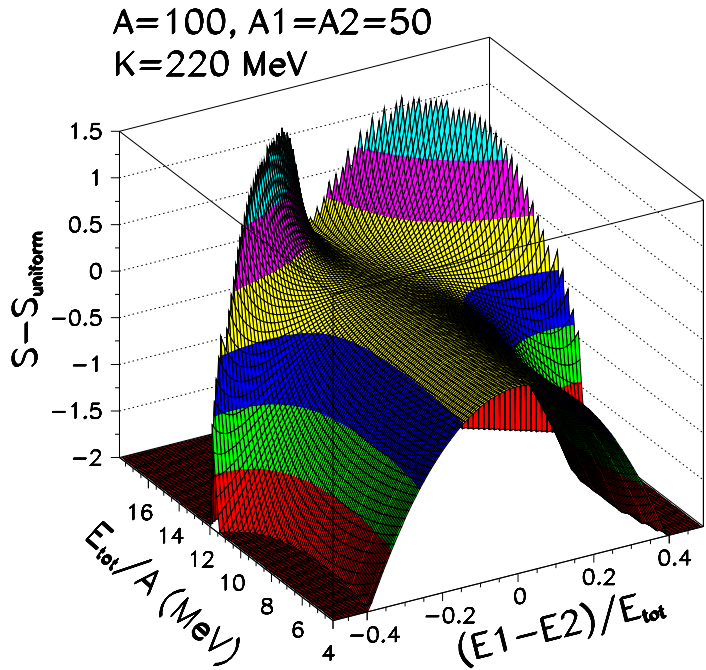}
\caption{(Color online) Reduced two-phase configuration entropy surface for a configuration of two equal-A subsystems with differing split of the available excitation energy $E^*$ between these phases.}
\label{fig_3}
\end{figure}

A further insight into the (volume) boiling phenomenon is provided in Fig.~\ref{fig_3} where the reduced two-phase configuration entropy $S_{config}-S_{uniform}$ surface is shown as a function of total excitation energy and its possible division between the two equal parts of the system. As seen in this figure, at low excitations entropy favors uniform configurations where the two parts have equal excitation energies. With increasing excitation energy, fluctuations in the excitation energy distribution grow to the point where a uniform distribution no longer provides a fair description of the system and, accordingly, the configuration temperature no longer provides an adequate representation of the system temperature. With a further increase of excitation energy, the entropy favors asymmetric split of the excitation energy between the two parts. In fact, the entropy would grow indefinitely with an indefinite expansion of one part of the system at the expense of the remainder until that part leaves the system or in other words, boils off. Note that in Fig.~\ref{fig_3} the mass numbers in the two parts of the model system are kept constant but the volumes occupied by these parts are free to adjust so as to maximize the configuration entropy for a given split of excitation energy.

The results show that in quantitative terms the model volume boiling temperature of over 11 MeV is substantially higher than the limiting temperatures observed in experimental studies of caloric curves \cite{natowitz} making it unlikely for the volume boiling to be responsible for the observed plateaus on these curves. On the other hand, as demonstrated in the following section, in finite systems, the more vulnerable surface domain begins boiling off at lower temperatures consistent with the observed plateaus on caloric curves.

\section{Surface boiling in finite systems}
\label{section_surfaceboiling}

The model calculations for a finite system were performed for a system of 100 nucleons. The folding width parameter $\mu$ was 1.4fm.
The system density distribution was parameterized in terms of error function
\cite{toke-swiatecki} as
\begin{equation}
{\rho\over\rho_o}=C(R_{half},d)[1-erf({r-R_{half}\over \sqrt{2}d})],
\label{eq_profile}
\end{equation}
\noindent where $R_{half}$ and $d$ are the half-density radius and the S\"ussmann surface width, respectively, and $C(R_{half},d)$ is a normalization factor assuring the desired number of nucleons in the system (here, A=100).

\begin{figure}
\includegraphics [width=8cm ]{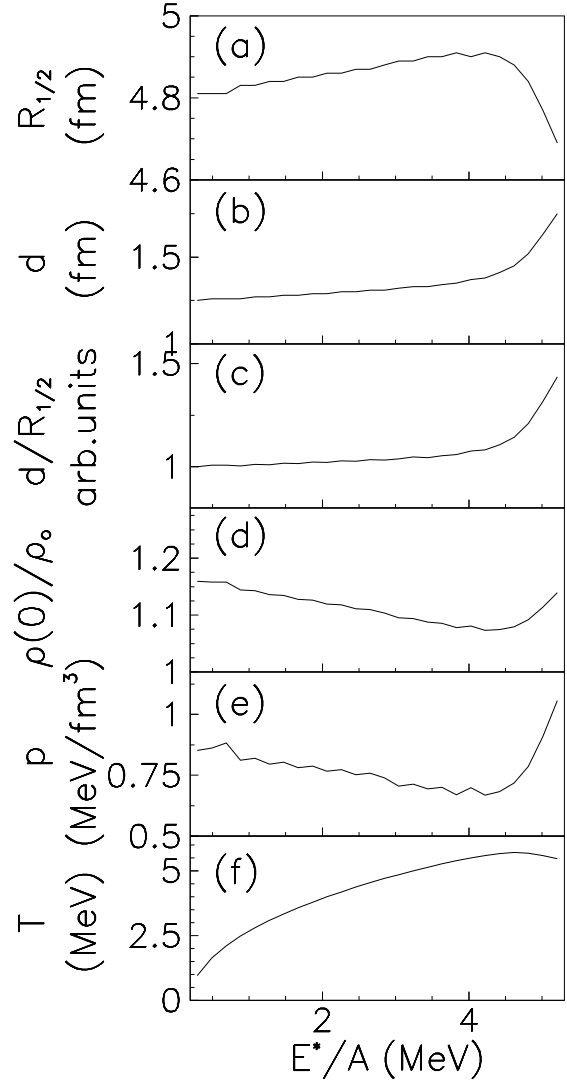}
\caption{Evolution of finite system parameters with excitation energy (See text).}
\label{fig_4}
\end{figure}

The results of the calculations are displayed in six panels on Fig.~\ref{fig_4}. The various panels illustrate the evolution of the system parameters with the excitation energy per nucleon. As seen in this figure, the system expands (panel a) as the excitation energy is increased up to 4 MeV per nucleon and so does the surface width (panel b) and the relative surface width (panel c). Accordingly, the bulk matter density in the center of the system decreases (panel e). The central pressure is seen to decrease (panel e) due to the reduction in surface tension. Importantly, the caloric curve (panel f) features a maximum followed with a domain of negative heat capacity. The negative heat capacity is seen to set in around an excitation energy per nucleon of approximately 5 MeV. It signals onset of an instability where a part of the surface domain may draw excitation energy from neighboring parts and while doing so expands and cools down. As a result, it will now draw even more energy from hotter neighbors and cool down even further until it separates from the system in a process that can be identified as surface boiling. The situation here is very much analogous to the case of volume boiling, except that the surface boiling sets in at much lower temperatures consistent with weaker bonding in the surface domain.

The loss of monotonicity at around $E*/A$=4.5 MeV seen in panels a-e of Fig.~\ref{fig_4} is due to the fact that the rather arbitrarily chosen two-phase configuration entropy ceases to provide a good approximation for the system entropy as the system approaches the boiling point. Like in the case of infinite matter, also here the zero-pressure trajectory of metastability ends at a point on the energy scale, where no profile secures maximum of entropy allowing the surface diffuseness of parts of the system grow indefinitely on their own.

\section{summary}
\label{section_summary}

The modeling of the behavior of hot bulk matter and hot finite nuclei has revealed that both kinds of systems are subject to boiling off of matter when brought to excitation energy per nucleon exceeding certain critical (not to be confused with critical point) value. In realistic finite systems the surface boiling sets in at considerably lower excitation energies than would the volume boiling of the bulk matter, which should prevent the system from ever experiencing volume boiling. On the other hand, the surface boiling, occurring in model calculations at temperatures around 5 MeV provides a natural explanation for the observed plateaus on caloric curves and observed limiting excitation energies that can be equilibrated in nuclear systems. The surface boiling appears unique to small systems where the surface domain plays relatively important role. Importantly, it appears impossible to avoid it as one ``pumps'' more and more excitation energy into the system. It simply is not possible to generate excitation at a rate slow enough to allow the excess to be carried away via statistical particle evaporation.

It appears far from obvious what form the boiled off matter may take, and specifically, whether it will cluster occasionally into intermediate-mass fragments. However, violent departure of parts of the surface domain may well give rise to shape fluctuation leading to Coulomb fragmentation \cite{toke_signatures}.

Consistent with the goals set, the present study employs a number of approximations and simplifications, which may be dropped in follow-up studies. For example, the developed general formalism can be readily adapted for handling iso-asymmetric systems allowing one to study isospin effects in boiling off of surface matter. Also, the quantum effects studied in Ref.~\cite{toke_quantum} should reduce the density of the bulk matter, depending on the size of the system. One may then wish to investigate the influence of such quantum effects on the boiling temperatures of systems of various sizes.

\begin{acknowledgments}
This work was supported by the U.S. Department of Energy grant No.
DE-FG02-88ER40414.
\end{acknowledgments}

%\bibliography{references}

\begin{thebibliography}{16}
\expandafter\ifx\csname natexlab\endcsname\relax\def\natexlab#1{#1}\fi
\expandafter\ifx\csname bibnamefont\endcsname\relax
  \def\bibnamefont#1{#1}\fi
\expandafter\ifx\csname bibfnamefont\endcsname\relax
  \def\bibfnamefont#1{#1}\fi
\expandafter\ifx\csname citenamefont\endcsname\relax
  \def\citenamefont#1{#1}\fi
\expandafter\ifx\csname url\endcsname\relax
  \def\url#1{\texttt{#1}}\fi
\expandafter\ifx\csname urlprefix\endcsname\relax\def\urlprefix{URL }\fi
\providecommand{\bibinfo}[2]{#2}
\providecommand{\eprint}[2][]{\url{#2}}

\bibitem[{\citenamefont{M{\"u}ller and Serot}(1995)}]{muller_serot}
\bibinfo{author}{\bibfnamefont{H.}~\bibnamefont{M{\"u}ller}} \bibnamefont{and}
  \bibinfo{author}{\bibfnamefont{B.}~\bibnamefont{Serot}},
  \bibinfo{journal}{Phys. Rev. C} \textbf{\bibinfo{volume}{52}},
  \bibinfo{pages}{2072} (\bibinfo{year}{1995}).

\bibitem[{\citenamefont{Natowitz et~al.}(2002)}]{natowitz}
\bibinfo{author}{\bibfnamefont{J.}~\bibnamefont{Natowitz}}
  \bibnamefont{et~al.}, \bibinfo{journal}{Phys. Rev. C}
  \textbf{\bibinfo{volume}{65}}, \bibinfo{pages}{34618} (\bibinfo{year}{2002}).

\bibitem[{\citenamefont{Shlomo and Kolomietz}(2005)}]{shlomo_kolomietz}
\bibinfo{author}{\bibfnamefont{S.}~\bibnamefont{Shlomo}} \bibnamefont{and}
  \bibinfo{author}{\bibfnamefont{V.}~\bibnamefont{Kolomietz}},
  \bibinfo{journal}{Rep> Prog. Phys.} \textbf{\bibinfo{volume}{68}},
  \bibinfo{pages}{1} (\bibinfo{year}{2005}).

\bibitem[{\citenamefont{Lattimer and Ravenhall}(1978)}]{lattimer_ravenhall}
\bibinfo{author}{\bibfnamefont{J.}~\bibnamefont{Lattimer}} \bibnamefont{and}
  \bibinfo{author}{\bibfnamefont{D.}~\bibnamefont{Ravenhall}},
  \bibinfo{journal}{Astr. Journ.} \textbf{\bibinfo{volume}{223}},
  \bibinfo{pages}{314} (\bibinfo{year}{1978}).

\bibitem[{\citenamefont{Barranco and Buchler}(1980)}]{barranco_buchler}
\bibinfo{author}{\bibfnamefont{M.}~\bibnamefont{Barranco}} \bibnamefont{and}
  \bibinfo{author}{\bibfnamefont{J.}~\bibnamefont{Buchler}},
  \bibinfo{journal}{Phys. Rev. C} \textbf{\bibinfo{volume}{22}},
  \bibinfo{pages}{1729} (\bibinfo{year}{1980}).

\bibitem[{\citenamefont{T{\~o}ke and Swiatecki}(1981)}]{toke-swiatecki}
\bibinfo{author}{\bibfnamefont{J.}~\bibnamefont{T{\~o}ke}} \bibnamefont{and}
  \bibinfo{author}{\bibfnamefont{W.~J.} \bibnamefont{Swiatecki}},
  \bibinfo{journal}{Nucl. Phys. A} \textbf{\bibinfo{volume}{372}},
  \bibinfo{pages}{141} (\bibinfo{year}{1981}).

\bibitem[{\citenamefont{T{\~o}ke and Schr{\"o}der}(1999)}]{toke_surface}
\bibinfo{author}{\bibfnamefont{J.}~\bibnamefont{T{\~o}ke}} \bibnamefont{and}
  \bibinfo{author}{\bibfnamefont{W.~U.} \bibnamefont{Schr{\"o}der}},
  \bibinfo{journal}{Phys. Rev. Lett.} \textbf{\bibinfo{volume}{82}},
  \bibinfo{pages}{5008} (\bibinfo{year}{1999}).

\bibitem[{\citenamefont{T{\~o}ke
  et~al.}(2003{\natexlab{a}})\citenamefont{T{\~o}ke, Lu, and
  Schr{\"o}der}}]{toke_surfentr}
\bibinfo{author}{\bibfnamefont{J.}~\bibnamefont{T{\~o}ke}},
  \bibinfo{author}{\bibfnamefont{J.}~\bibnamefont{Lu}}, \bibnamefont{and}
  \bibinfo{author}{\bibfnamefont{W.~U.} \bibnamefont{Schr{\"o}der}},
  \bibinfo{journal}{Phys. Rev. C} \textbf{\bibinfo{volume}{67}},
  \bibinfo{pages}{034609} (\bibinfo{year}{2003}{\natexlab{a}}).

\bibitem[{\citenamefont{T{\~o}ke and Schr{\"o}der}(2002)}]{toke_quantum}
\bibinfo{author}{\bibfnamefont{J.}~\bibnamefont{T{\~o}ke}} \bibnamefont{and}
  \bibinfo{author}{\bibfnamefont{W.}~\bibnamefont{Schr{\"o}der}},
  \bibinfo{journal}{Phys. Rev. C} \textbf{\bibinfo{volume}{65}},
  \bibinfo{pages}{044319} (\bibinfo{year}{2002}).

\bibitem[{\citenamefont{T{\~o}ke et~al.}(2005)\citenamefont{T{\~o}ke,
  Pie\'nkowski, Sobotka, Houck, and Schr{\"o}der}}]{toke_retard}
\bibinfo{author}{\bibfnamefont{J.}~\bibnamefont{T{\~o}ke}},
  \bibinfo{author}{\bibfnamefont{L.}~\bibnamefont{Pie\'nkowski}},
  \bibinfo{author}{\bibfnamefont{L.}~\bibnamefont{Sobotka}},
  \bibinfo{author}{\bibfnamefont{M.}~\bibnamefont{Houck}}, \bibnamefont{and}
  \bibinfo{author}{\bibfnamefont{W.~U.} \bibnamefont{Schr{\"o}der}},
  \bibinfo{journal}{Phys. Rev. C} \textbf{\bibinfo{volume}{72}},
  \bibinfo{pages}{031601} (\bibinfo{year}{2005}).

\bibitem[{\citenamefont{T{\~o}ke
  et~al.}(2003{\natexlab{b}})\citenamefont{T{\~o}ke, Lu, and
  Schr{\"o}der}}]{toke_box}
\bibinfo{author}{\bibfnamefont{J.}~\bibnamefont{T{\~o}ke}},
  \bibinfo{author}{\bibfnamefont{J.}~\bibnamefont{Lu}}, \bibnamefont{and}
  \bibinfo{author}{\bibfnamefont{W.~U.} \bibnamefont{Schr{\"o}der}},
  \bibinfo{journal}{Phys. Rev. C} \textbf{\bibinfo{volume}{67}},
  \bibinfo{pages}{044307} (\bibinfo{year}{2003}{\natexlab{b}}).

\bibitem[{\citenamefont{T{\~o}ke and Schr{\"o}der}(2006)}]{toke_unified}
\bibinfo{author}{\bibfnamefont{J.}~\bibnamefont{T{\~o}ke}} \bibnamefont{and}
  \bibinfo{author}{\bibfnamefont{W.~U.} \bibnamefont{Schr{\"o}der}},
  \emph{\bibinfo{title}{Proc. IWM2005 Int. Worksh. on Multifragmentation and
  Rel. Topics}} (\bibinfo{publisher}{Societ\`a Italiana di Fisica},
  \bibinfo{address}{Bologna, Italy}, \bibinfo{year}{2006}), p.
  \bibinfo{pages}{379}.

\bibitem[{\citenamefont{T{\~o}ke and Schr{\"o}der}(2009)}]{toke_signatures}
\bibinfo{author}{\bibfnamefont{J.}~\bibnamefont{T{\~o}ke}} \bibnamefont{and}
  \bibinfo{author}{\bibfnamefont{W.}~\bibnamefont{Schr{\"o}der}},
  \bibinfo{journal}{Phys. Rev. C} \textbf{\bibinfo{volume}{79}},
  \bibinfo{pages}{064622} (\bibinfo{year}{2009}).

\bibitem[{\citenamefont{Sobotka et~al.}(2004)\citenamefont{Sobotka, Charity,
  T{\~o}ke, and Schr{\"o}der}}]{sobotka_mono}
\bibinfo{author}{\bibfnamefont{L.~G.} \bibnamefont{Sobotka}},
  \bibinfo{author}{\bibfnamefont{R.~J.} \bibnamefont{Charity}},
  \bibinfo{author}{\bibfnamefont{J.}~\bibnamefont{T{\~o}ke}}, \bibnamefont{and}
  \bibinfo{author}{\bibfnamefont{W.~U.} \bibnamefont{Schr{\"o}der}},
  \bibinfo{journal}{Phys. Rev. Lett.} \textbf{\bibinfo{volume}{93}},
  \bibinfo{pages}{132702} (\bibinfo{year}{2004}).

\bibitem[{\citenamefont{Sobotka et~al.}(2006)\citenamefont{Sobotka, and
  Charity}}]{sobotka_2006}
  \bibinfo{author}{\bibfnamefont{L.~G.} \bibnamefont{Sobotka}},
  \bibnamefont{and} \bibinfo{author}{\bibfnamefont{R.~J.}
  \bibnamefont{Charity}}, \bibinfo{journal}{Phys. Rev. C}
  \textbf{\bibinfo{volume}{73}}, \bibinfo{pages}{014609}
  (\bibinfo{year}{2006}).

\bibitem[{\citenamefont{Hoel et~al.}(2007)\citenamefont{Hoel, Sobotka, and
  Charity}}]{sobotka_iso}
\bibinfo{author}{\bibfnamefont{C.}~\bibnamefont{Hoel}},
  \bibinfo{author}{\bibfnamefont{L.~G.} \bibnamefont{Sobotka}},
  \bibnamefont{and} \bibinfo{author}{\bibfnamefont{R.~J.}
  \bibnamefont{Charity}}, \bibinfo{journal}{Phys. Rev. C}
  \textbf{\bibinfo{volume}{75}}, \bibinfo{pages}{017601}
  (\bibinfo{year}{2007}).

\bibitem[{\citenamefont{Friedman}(1988)}]{eesm}
\bibinfo{author}{\bibfnamefont{W.~A.} \bibnamefont{Friedman}},
  \bibinfo{journal}{Phys. Rev. Lett.} \textbf{\bibinfo{volume}{60}},
  \bibinfo{pages}{2125} (\bibinfo{year}{1988}).

\bibitem[{\citenamefont{Hugenholtz and {Van Hove}}(1958)}]{hvh}
\bibinfo{author}{\bibfnamefont{N.}~\bibnamefont{Hugenholtz}} \bibnamefont{and}
  \bibinfo{author}{\bibfnamefont{L.}~\bibnamefont{{Van Hove}}},
  \bibinfo{journal}{Physica} \textbf{\bibinfo{volume}{24}},
  \bibinfo{pages}{363} (\bibinfo{year}{1958}).

\end{thebibliography}

\end{document}